# Generalized Principle of limiting 4-dimensional symmetry. Relativistic length expansion in accelerated system revisited.

Jaykov Foukzon
Israel Institute of Technology

jaykovfoukzon@list.ru
S.A.Podosenov

All-Russian Scientific-Research Institute
for Optical and Physical Measurements,
Moscow 119361, Russia

**Abstract:** In this article, Generalized Principle of "limiting 4-dimensional symmetry": *The laws of physics in non-inertial frames must display the 4-dimensional symmetry of the Generalized Lorentz-Poincare group in the limit of zero acceleration,* is proposed. Classical solution of the relativistic length expansion in general accelerated system revisited.

## I. Introduction

In his famous 1905 paper, Einstein proposed that all physical theories should satisfy the (now well-known) two postulates of special relativity:

**Axiom 1.** *The physical laws of nature and the results of all experiments are independent of the particular inertial frame of the observer (in which the experiment is performed); and*

**Axiom 2.** *The speed of light is independent of the motion of the source.*

**Remark** *Consider the Galilean group:*

$$x = x' - \mathbf{v}t'; t = t', \qquad (1.1)$$

or in equivalent infinitesimal form

$$dx = dx' - \mathbf{v}dt'; dt = dt', \qquad (1.2)$$

From the above two postulates *Galilean group (0.1) was changed by Lorentz group:*

$$x = \gamma(\mathbf{v})\{x' - \mathbf{v}t'\},$$
$$t = \gamma(\mathbf{v})\left\{t' - \frac{\mathbf{v}}{c^2}x'\right\},$$
$$y = y', z = z', \tag{1.3}$$
$$\gamma(\mathbf{v}) = \frac{1}{\sqrt{1 - \frac{\mathbf{v}^2}{c^2}}}.$$

or in equivalent infinitesimal form

$$dx = \gamma(\mathbf{v})\{dx' - \mathbf{v}dt'\},$$
$$dt = \gamma(\mathbf{v})\left\{dt' - \frac{\mathbf{v}}{c^2}dx'\right\}, \tag{1.4}$$
$$dy = dy', dz = dz'.$$

**Axiom** *3. The correct implementation of postulates 1 and 2 is to represent time as a fourth coordinate and constrain the relationship between components so as to satisfy the natural invariance induced by the Lorentz group (of electromagnetism). As is well known, this procedure leads to the concept of Minkowski space $M_4$.*

**Notation** *1.The third postulate was made by Minkowski [1],[2], a well-known mathematician, and was embraced by many.*

**Remark** *In other words, we postulate that in the whole space there is a physical frame of reference (FR) called an inertial (Galilee) one in which the interval between events of this space is written as*

$$ds'^2 = c^2 dt'^2 - dx'^2 - dy'^2 - dz'^2. \tag{1.5}$$

*In general, in the Minkowski space-time any FR in which the interval (1.1) has the general form*

$$ds^2 = g_{ik}dx^i dx^k; i,k = 0,1,2,3 \tag{1.6}$$

*and which satisfies the allowance conditions $g_{00} > 0; g_{\alpha\beta}dx^\alpha dx^\beta < 0; \alpha,\beta = 1,2,3$ is allowed.*

**Theorem** *The curvature tensor becomes zero identically: $\mathbf{R}_{iklm} = 0$ in any FR of the Minkowski space, including non-inertial (accelerated) one. The transformations connecting FR's of the real moving bodies (particles of matter) with the initial inertial Galilee FR (1.5) can be either linear or non-linear ones.The first ones (which are always non-orthogonal ones for the real bodies and, therefore, not coinciding with the Lorentz transformations!) form a so-called generalized inertial FR (with non-orthogonal axes $t,x$) the metrics of which has the off-diagonal part $g_{0i} = 0$, the second ones give in turn the non-inertial (accelerated) FR's (also with a nondiagonal metrics). The particular case of the generalized inertial FR is the FR connected with the inertial one (1.5) by the classic Galilee transformation: $x' = x + \mathbf{v}t; t = t'$, and corresponding to rotation of the axis $t'$ with fixed orientation of the axis $x'$.Let us pass from Galilee's coordinates $(t',x',y',z')$ with the metric (1.5) to*

*coordinates $(x,y,z,t)$ by arbitrary linear transformation. This transformation is equivalent up to a space axis rotation to a transformation in plane $t',x'$ :*

$$x' = ax + bt; t' = qx + pt; y' = y; z' = z. \qquad (1.7)$$

*Substituting (1.7) in (1.5), in coordinates $(x,y,z,t)$ the metric gets the form:*

$$ds^2 = c^2 g_{00} dt^2 + 2c g_{01} dt dx + g_{11} dx^2 - dy^2 - dz^2 \qquad (1.8)$$

*where $g_{00} = p^2 - b^2/c^2; g_{01} = c(pq - ab/c^2); g_{11} = c^2 q^2 - a^2$. The transformation (1.7) describes the rotation of the axes $x,t$ in the plane $x',t'$, with after the rotation the axis $x$ can be not orthogonal to the axis $t$, i.e. $x$ and $t$ rotate on angles, which may differ. The metric (1.8) gives a generalized inertial frame of reference in the SRT. The Lorentz transformations are a particular case of the general linear transformations 3, corresponding to the choice $g_{00} = 1, g_{01} = 0, g_{11} = -1$ in (1.8). Hence, the metric (1.8), in contrast to (1.5), is not forminvariant with respect to the Lorentz transformations. Let us consider a rotation of axis $t$ without changing of the $x$ orientation as particular case of the transformation (1.7). It is the classic Galilee transformation:*

$$x' = x + \mathbf{v}_0 t; t' = t, \mathbf{v} = \mathbf{const}. \qquad (1.9)$$

*corresponding to the choice of parameters in (1.7) as $p = a = 1; q = 0; b = \mathbf{v}_0$. Thus, the metric (1.8) get the form:*

$$ds^2 = c^2\left(1 - \frac{\mathbf{v}_0^2}{c^2}\right) dt^2 - 2\mathbf{v}_0 dt dx - dx^2 - dy^2 - dz^2. \qquad (1.10)$$

**Theorem** *1.1. The metrics of the inertial (Galilee) FR (1.5) is forminvariant with respect to the classic Lorentz-Poincare transformation group $L_m^i$. The metrics of the generalized inertial FR is that with respect to the so-called generalized inertial Lorentz-Poincare group $\hat{L}_m^i$ connected with the classic one by the relation:*

$$\hat{L}_k^n x^k = [\Lambda_i^n L_m^i (\Lambda^{-1})_k^m] x^k, \qquad (1.11)$$

*where $\Lambda_k^i$ is the matrix of the linear (non-orthogonal) transformations forming the generalized inertial FR*

$$x^i = \Lambda_k^i x'^k. \qquad (1.12)$$

The transformations (1.11) are orthogonal but connect the number of non-orthogonal FR's with the same nondiagonal metrics. In the particular case of the Galilee transformation with metric (1.10) the group of transformations, keeping the metric (1.10) forminvariant, takes the form:

$$x = \gamma(\mathbf{v})\left\{\left(1 + \frac{\mathbf{v}_0 \mathbf{v}}{c^2}\right) x' + \left(1 - \frac{\mathbf{v}_0^2}{c^2}\right) \mathbf{v} t'\right\},$$
$$t = \gamma(\mathbf{v})\left\{\frac{\mathbf{v}^2}{c^2} x' + \left(1 - \frac{\mathbf{v}_0 \mathbf{v}}{c^2}\right) t'\right\}. \qquad (1.13)$$

At $u = 0$ transformations (1.13) coincides with the Lorentz transformation, naturally.

Almost all physical frames of reference in the universe are, strictly speaking, non-inertial because of the long range action of the gravitational force. Thus it is desirable that the laws of physics and the universal and fundamental constants are understood or known not only in inertial frames but also in non-inertial frames.So it is natural and necessary to require that the laws of physics in non-inertial frames must display the 4-dimensional symmetry of the generalized Lorentz-Poincare group in the limit of zero acceleration.Such a requirement is postulated as the principle of Generalized limiting 4-dimensional symmetry

**Axiom** 4.(**Generalized Principle of limiting 4-dimensional symmetry**)

The laws of physics in non-inertial frames must display the 4-dimensional symmetry of the Generalized Lorentz-Poincare group in the limit of zero acceleration.

In particular from Axiom 4 we obtain

**Hsu's Principle of limiting 4-dimensional symmetry** [3]:

The laws of physics in non-inertial *orthogonal* frames must display the 4-dimensional symmetry of the Lorentz and Poincarr'e groups in the limit of zero acceleration.

Let us consider transformations for the constant-linear-acceleration frame $F_w(x)$ and an inertial frame $F_I(x_I = x')$

$$t' = \gamma\beta\left(x + \frac{1}{w_0\gamma_0^2}\right) + a_0, x' = \gamma\left(x + \frac{1}{w_0\gamma_0^2}\right) + b_0$$
$$y' = y, z' = z,$$
$$\beta = w_0 t + \beta_0, \gamma = \frac{1}{\sqrt{1-\beta^2}}, \gamma_0 = \frac{1}{\sqrt{1-\beta_0^2}},$$
$$a_0 = -\frac{\beta_0}{w_0\gamma_0}, b_0 = \frac{1}{w_0\gamma_0}.$$
(1.14)

where the velocity $\beta = \beta(t)$ is a linear function time $t$. The result (1.14) is called the Wu transformation. It reduces to the Möller transformation when $\beta_0 = 0$, provided a change of time variable ($t = (1/w_0)\tanh(w_0 t^*)$) is made [3].

Furthermore, in the limit of zero acceleration,$w_0 \to 0$, the Wu transformation (1.14) reduces to the 4-dimensional transformations (which form the Lorentz group),

$$t' = \gamma_0(t + \beta_0 x), x' = \gamma_0(x + \beta_0 t)$$
$$y' = y, z' = z.$$
(1.15)

Thus, limiting 4-dimensional symmetry of the Lorentz and Poincarr'e invariance is satisfied. The differential form of the Wu transformation (1.14) for constant-linear-acceleration is

$$dt' = \gamma(W_c dt + \beta dx), dx' = \gamma(dx + \beta W_c dt)$$
$$dy' = dy, dz' = dz, \quad (1.16)$$
$$W_c = \gamma^2(\gamma_0^{-2} + w_0 x).$$

Based on the differential from in (1.16), we can consider the generalization of the Wu transformation (1.15) to a more general non-inertial frame $F(x; w(t))$ moving with an arbitrary velocity $\beta(t)$ or arbitrary acceleration $w(t)$ along the $x$-axis

$$\beta(t) = \beta_1(t) + \beta_0, w(t) = \frac{d\beta(t)}{dt} = \frac{d\beta_1(t)}{dt}. \quad (1.17)$$
$$\beta(0) = \beta_0, w(0) = w_0.$$

A simple and general spacetime transformation for GLA frames is [3]:

$$t' = \gamma\beta\left(x + \frac{1}{w(t)\gamma_0^2}\right) - \frac{\beta_0}{w_0\gamma_0},$$
$$x' = \gamma\left(x + \frac{1}{w(t)\gamma_0^2}\right) + \frac{1}{w_0\gamma_0}, \quad (1.18)$$
$$y' = y, z' = z.$$

where the two constants of integration $a_0$ and $b_0$ are determined by the limiting 4-dimensional symmetry as $w(0) = w_0 \to 0$.

From the transformation (1.18), we can obtain a simple transformations for the differentials $dx^\mu$ and $dx'^\mu$:

$$dt' = \gamma(t)[W_1(t)dt + \beta(t)dx], dx' = \gamma(t)[dx + \beta(t)W_2(t)dt]$$
$$dy' = dy, dz' = dz.$$
$$W_1(t) = \gamma^2(t)\left(w(t)x + \frac{1}{\gamma_0^2}\right) - \frac{\beta(t)J_e(t)}{w^2(t)\gamma_0^2} > 0, \quad (1.19)$$
$$W_2(t) = \gamma^2(t)\left(w(t)x + \frac{1}{\gamma_0^2}\right) - \frac{J_e(t)}{w^2(t)\beta(t)\gamma_0^2} > 0.$$

The invariant infinitesimal interval $ds^2$ in GLA frames can be obtained from (1.19).

$$ds^2 = dt'^2 - dx'^2 - dy'^2 - dz'^2 = g_{\mu\nu}dx^\mu dx^\nu =$$
$$W^2(t,x)dt^2 - 2U(t)dtdx - dx^2 - dy^2 - dz^2, \quad (1.20)$$
$$U(t) = \frac{J_e(t)}{w^2(t)\gamma_0^2}, J_e(t) = \frac{dw(t)}{dt}.$$

When the jerk $J_e(t)$ vanishes, we have $w(t) = w_0$ and one can see that the transformation (1.20) reduces to the Wu transformation (1.14) for a constant-linear-acceleration frame $F_c(x)$, in which the time axis is everywhere orthogonal to the spatial coordinate curves.

# II. Generalized Principle of limiting 4-dimensional symmetry

Let us considered general (acceleration) transformations between the two relativistic frames one of which inertial frame $K' = K'(t,x',y',z')$ will be considered to be at "rest", while another one accelerated frame $K = K(t,x,y,z)$ will move with respect to the first one by the law:

$$\begin{aligned} x' &= F_1(t,x), \\ t' &= F_2(t,x), \\ y' &= y, z' &= z. \end{aligned} \qquad (2.1)$$

or in equivalent infinitesimal form:

$$\begin{aligned} dx' &= DF_1(t,x) = \frac{\partial F_1}{\partial x}dx + \frac{\partial F_1}{\partial t}dt, \\ dt' &= DF_2(t,x) = \frac{\partial F_2}{\partial x}dx + \frac{\partial F_2}{\partial t}dt, \\ dy' &= dy, dz' = dz. \end{aligned} \qquad (2.2)$$

Thus

$$\begin{aligned} dt'^2 &= \left(\frac{\partial F_2}{\partial x}dx + \frac{\partial F_2}{\partial t}dt\right)^2 = F_{2x}^2 dx^2 + 2F_{2x}F_{2t}dtdx + F_{2t}^2 dt^2, \\ dx'^2 &= \left(\frac{\partial F_1}{\partial x}dx + \frac{\partial F_1}{\partial t}dt\right)^2 = F_{1x}^2 dx^2 + 2F_{1x}F_{1t}dtdx + F_{1t}^2 dt^2 \end{aligned} \qquad (2.3)$$

By substituting Eqs.(2.3) into $ds^2 = c^2 dt'^2 - dx'^2 - dy'^2 - dz'^2$ we obtain

$$\begin{aligned} ds^2 &= c^2 F_{2x}^2 dx^2 + 2c^2 F_{2x}F_{2t}dtdx + c^2 F_{2t}^2 dt^2 - F_{1x}^2 dx^2 - 2F_{1x}F_{1t}dtdx - F_{1t}^2 dt^2 - \\ &\quad -dy'^2 - dz'^2 = \\ &= c^2\left(F_{2t}^2 - \frac{F_{1t}^2}{c^2}\right)dt^2 + 2(c^2 F_{2x}F_{2t} - F_{1x}F_{1t})dtdx + (c^2 F_{2x}^2 - F_{1x}^2)dx^2 - dy'^2 - dz'^2. \end{aligned} \qquad (2.4)$$

In the limit of zero acceleration, transformations Eqs.(2.2) becomes to the next form

$$dx' = a(x)dx + b(x)dt; dt' = q(x)dx + p(x)dt; y' = y; z' = z. \qquad (2.5)$$

and metric (2.4) gets the form:

$$ds^2 = c^2 g_{00}(x)dt^2 + 2c g_{01}(x)dtdx + g_{11}(x)dx^2 - dy^2 - dz^2 \qquad (2.6)$$

where

$$g_{00} = p^2(x) - b^2(x)/c^2; g_{01} = c(p(x)q(x) - a(x)b(x)/c^2); g_{11} = c^2q^2(x) - a^2(x). \quad (2.7)$$

**Theorem** *The metrics (2.4) of the general noninertial FR in the limit of zero accelerationis is forminvariant that with respect to the so-called generalized inertial Lorentz-Poincare group $\hat{L}_m^i(x)$ connected with the classic Lorentz-Poincare group $L_m^i$ one by the relation:*

$$\hat{L}_k^n(x)x^k = [\Lambda_i^n(x)L_m^i(\Lambda^{-1}(x))_k^m]x^k, \quad (2.8)$$

*where $\Lambda_k^i(x)$ is the matrix of the linear (non-orthogonal) transformations*

$$x^i = \Lambda_k^i(x)x'^k. \quad (2.9)$$

*forming the generalized inertial FR (2.6).*

# III. Generalized Principle of limiting 4-dimensional symmetry for the case of the uniformly accelerated frames of reference

Let us considered (acceleration) transformations between the two relativistic frames one of which inertial frame $K' = K'(t, x', y', z')$ will be considered to be at "rest", while another one uniformly accelerated frame $K = K(t, x, y, z)$ will move with respect to the first one by the law:

$$x' = x + \int_0^t \mathbf{v}(\tau)d\tau, t = t', \quad (3.1)$$

or in equivalent infinitesimal forms

$$dx' = dx + \mathbf{v}(t)dt, \mathbf{v}(t) < c; t' = t, \quad (3.2)$$

Thus general metric (2.4) gets the form:

$$ds^2 = c^2\left(1 - \frac{\mathbf{v}^2(t)}{c^2}\right)dt^2 - 2\mathbf{v}(t)dtdx - dx^2 - dy'^2 - dz'^2. \quad (3.3)$$

In the limit of zero acceleration $\lim_{t \to T} \vec{a}(t) \to 0,$ we have $\mathbf{v}(t) = \mathbf{v}(T) = \mathbf{v}_T, t \geqslant T$ and transformations (3.1) becomes to the form (1.9) and metric (3.3) gets the form: (see Eq.1.10)

$$ds^2 = c^2\left(1 - \frac{\mathbf{v}_T^2}{c^2}\right)dt^2 - 2\mathbf{v}_T dtdx - dx^2 - dy'^2 - dz'^2. \tag{3.4}$$

**Theorem** 3.1. *The metrics (3.3) of the uniformly accelerated noninertial FR in the limit of zero accelerationis is forminvariant that with respect to the so-called generalized inertial Lorentz-Poincare group $\hat{L}_m^i$ connected with the classic Lorentz-Poincare group $L_m^i$ one by the relation:*

$$\hat{L}_k^n x^k = [\Lambda_i^n L_m^i (\Lambda^{-1})_k^m] x^k, \tag{3.5}$$

*where $\Lambda_k^i$ is the matrix of the linear (non-orthogonal) transformations*

$$x^i = \Lambda_k^i x'^k. \tag{3.6}$$

*forming the generalized inertial FR (3.4).*

**Remark** *Relativistic motion with uniformly acceleration is a motion under the influence of a uniformly force $\vec{f}(t)$, that is uniformly in value and direction $\vec{r}$. According to the equations of relativistic motion we have*

$$\frac{d}{dt}\left(\frac{\vec{v}(t)}{\sqrt{1 - \frac{v^2(t)}{c^2}}}\right) = \frac{\vec{f}(t)}{m} = \vec{a}(t). \tag{3.7}$$

*Integrating equation (3.6) over time, we obtain*

$$\frac{\vec{v}(t)}{\sqrt{1 - \frac{v^2(t)}{c^2}}} = \vec{\beta}(t) + \vec{v}_0. \quad (a)$$

$$\vec{\beta}(t) = \int_0^t \vec{a}(\tau)d\tau. \quad (b) \tag{3.8}$$

*Setting the constant $v_0$ to zero, which corresponds to zero initial velocity, we find after squaring*

$$\frac{1}{1 - \frac{v^2(t)}{c^2}} = 1 + \frac{\beta^2(t)}{c^2}. \tag{3.9}$$

*Taking into account this expression in (3.7), we obtain*

$$\vec{v}(t) = \frac{d\vec{r}'}{dt} = \frac{\vec{\beta}(t)}{\sqrt{1 + \frac{\beta^2(t)}{c^2}}}. \tag{3.10}$$

*Integrating this equation, we find*

$$\vec{r}'(t) = \vec{r} + \int_0^t \frac{\beta(\tau)d\tau}{\sqrt{1 + \frac{\beta^2(\tau)}{c^2}}}. \qquad (3.11)$$

*In the limit of zero acceleration* $\lim_{t \to T} \vec{a}(t) \to 0$ *from Eq.(3.8.b) we obtain*

$$\lim_{t \to T} \beta(t) = \int_0^T \vec{a}(\tau)d\tau = \beta(T). \qquad (3.12)$$

Taking into account Eqs.(3.10),(3.12) we obtain

$$\mathbf{v}_T = \lim_{t \to T} \vec{\mathbf{v}}(t) = \vec{\mathbf{v}}(T)\frac{\beta(T)}{\sqrt{1 + \frac{\beta^2(T)}{c^2}}}. \qquad (3.13)$$

Thus in the limit of zero acceleration generalized Lorentz transformations (3.5) gets the form:

$$x = \gamma(\mathbf{V})\left\{\left(1 + \frac{\mathbf{v}_T\mathbf{V}}{c^2}\right)x' + \left(1 - \frac{\mathbf{v}_T^2}{c^2}\right)\mathbf{V}t'\right\},$$
$$t = \gamma(\mathbf{V})\left\{\frac{\mathbf{V}^2}{c^2}x' + \left(1 - \frac{\mathbf{v}_T\mathbf{V}}{c^2}\right)t'\right\}. \qquad (3.14)$$

# IV. Physical and coordinate values in SRT. Relativistic length expansion in accelerated system revisited.

As known [4] constructing the covariant SRT on a general non-inertial frame (1.6) one should exactly distinguish a coordinates (in some sense formal-mathematical) of a particles and *physical distance* (experimentally measurable) one. The latter is defined as the ratio of the *physical distance* $dl_{ph}$ and *physical time* $d\tau_{ph}$ :

$$ds^2 = c^2 d\tau_{\text{ph}}^2 - dl_{\text{ph}}^2,$$

$$dl_{\text{ph}}^2 = \left(-g_{\alpha\beta} + \frac{g_{0\alpha}g_{0\beta}}{g_{00}}\right)dx^\alpha dx^\beta, \tag{4.1}$$

$$d\tau_{\text{ph}} = \sqrt{g_{00}}\, dt + \frac{g_{0\alpha}dx^\alpha}{c\sqrt{g_{00}}}.$$

Let us consider measurement, in a general non-inertial reference frame (1.6), of the *physical length* $l_{\text{ph}}$ of a rod. We first determine the method for measuring the length of a moving *infinite small rod*. Consider an observer in the non-inertial reference system, who records the ends of the rod, $\mathbf{X}_1(\tau_{\text{ph}}(t))$ and $\mathbf{X}_2(\tau_{\text{ph}}(t))$, at the same moment of *physical* time $\tau_{\text{ph}}(t)$, i.e.

$$d\tau_{\text{ph}}(t) = \sqrt{g_{00}}\, dt + \frac{g_{0\alpha}dx^\alpha(t)}{c\sqrt{g_{00}}} = 0. \tag{4.2}$$

Hence using Eq.(4.2) one obtain

$$\sqrt{g_{00}} + \frac{g_{0\alpha}}{c\sqrt{g_{00}}}\frac{dx^\alpha}{dt} = 0 \tag{4.3}$$

Suppose that on the time interval $t \in [t_1, t_2]$ admissible solution $\tilde{x}^\alpha(t)$ of the Eq.(4.3) exist and the corresponding boundary conditions: $\tilde{x}^\alpha(t_1) = \tilde{x}_1^\alpha(t_1), \tilde{x}^\alpha(t_2) = \tilde{x}_2^\alpha(t_2)$ is satisfied. Under conditions (4.2)-(4.3) interval $ds^2 = c^2 d\tau_{\text{ph}}^2 - dl_{\text{ph}}^2$, changes from the interval with spatial part only:

$$ds^2 = -dl_{\text{ph}}^2,$$

$$dl_{\text{ph}} = \sqrt{\left(-g_{\alpha\beta}(\tilde{x}^\alpha(t),t) + \frac{g_{0\alpha}(\tilde{x}^\alpha(t),t)g_{0\beta}(\tilde{x}^\alpha(t),t)}{g_{00}(\tilde{x}^\alpha(t),t)}\right)\frac{d\tilde{x}^\alpha}{dt}\frac{d\tilde{x}^\beta}{dt}}\, dt, \tag{4.4}$$

$$l_{\text{ph}} = \text{sgn}(t_2 - t_1)\int_{t_1}^{t_2} \sqrt{\left(-g_{\alpha\beta}(\tilde{x}^\alpha(t),t) + \frac{g_{0\alpha}(\tilde{x}^\alpha(t),t)g_{0\beta}(\tilde{x}^\alpha(t),t)}{g_{00}(\tilde{x}^\alpha(t),t)}\right)\frac{d\tilde{x}^\alpha}{dt}\frac{d\tilde{x}^\beta}{dt}}\, dt.$$

**Example.** J.S. Bell's problem [7] (see also [8]). Its gist consists in the following. Two rockets **B** and **C** are set in motion (say, to the velocity $v(t) = \beta(t)c$ ) so that the distance between them may remains constant and equal to the starting one $L_0$ from the viewpoint of an external observer **A**. One can simpler imaging here that the observer A operates the flight of the moving-away rockets and has a radar, by means of which he controls the constancy of the distance between them. From the viewpoint of an external observer **A** rockets **C** and **B** moving by laws $x_C(t)$ and $x_B(t)$ :

$$x_C(t) = \frac{c^2}{a}\left(\sqrt{1 + \frac{a^2 t^2}{c^2}} - 1\right),$$

$$x_B(t) = L_0 + \frac{c^2}{a}\left(\sqrt{1 + \frac{a^2 t^2}{c^2}} - 1\right).$$

(4.5)

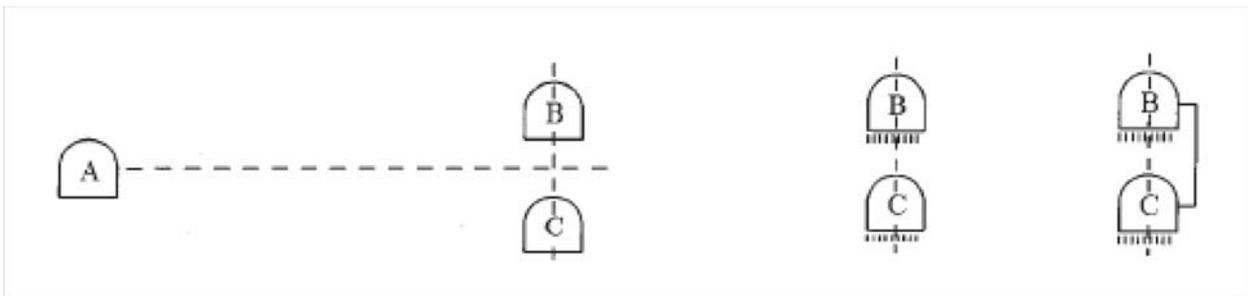

Pic.1.

In the case of motion with constant proper acceleration $a$ the interval (1.6) of the FR comuving both rockets takes the form (see [9] Eq.12.12):

$$ds^2 = \frac{c^2 dt^2}{1 + \frac{a^2 t^2}{c^2}} - \frac{2at\,dt\,dx}{\sqrt{1 + \frac{a^2 t^2}{c^2}}} - dx^2 - dy^2 - dz^2.$$

(4.6)

From Eqs.(4.3),(4.5) one obtain

$$\sqrt{g_{00}} + \frac{g_{01}}{c\sqrt{g_{00}}} \frac{dx^1}{dt} = 0,$$

$$\frac{dx^1}{dt} = -\frac{cg_{00}}{g_{01}} = \frac{c^2 \sqrt{1 + \frac{a^2 t^2}{c^2}}}{at\left(1 + \frac{a^2 t^2}{c^2}\right)} =$$

$$= \frac{c}{\frac{at}{c}\sqrt{1 + \frac{a^2 t^2}{c^2}}} \quad (4.7)$$

$$\frac{dx^1}{dt} = \frac{c}{\beta(t)\sqrt{1 + \beta^2(t)}},$$

$$\beta(t) = \frac{at}{c}.$$

Hence

$$x^1(t) = -\frac{c^2}{a} \ln\left|\frac{1 + \sqrt{1 + \beta^2(t)}}{\beta(t)}\right| + A. \quad (4.8)$$

Setting $t = t_2$ and take into account corresponding boundary condition $x^1(t) = x^1(t_2) = x_c(0) = 0$ one obtain:

$$-\frac{c^2}{a} \ln\left|\frac{1 + \sqrt{1 + \beta^2(t_2)}}{\beta(t_2)}\right| + A(t_2) = 0. \quad (4.9)$$

Thus

$$A(t_2) = \frac{c^2}{a} \ln\left|\frac{1 + \sqrt{1 + \beta^2(t_2)}}{\beta(t_2)}\right|. \quad (4.10)$$

Setting $t = t_1$ and take into account corresponding boundary

condition $x^1(t_1) = x_B(0) = L_0$ one obtain:

$$-\frac{c^2}{a} \ln\left|\frac{1 + \sqrt{1 + \beta^2(t_1)}}{\beta(t_1)}\right| + A(t_2) = L_0. \qquad (4.11)$$

Substitution Eq.(4.10) into Eq.(4.11) gives

$$-\frac{c^2}{a} \ln\left|\frac{1 + \sqrt{1 + \beta^2(t_1)}}{\beta(t_1)}\right| + \frac{c^2}{a} \ln\left|\frac{1 + \sqrt{1 + \beta^2(t_2)}}{\beta(t_2)}\right| = L_0.$$

Hence $\qquad (4.12)$

$$\frac{c^2}{a} \ln\left|\frac{\left(1 + \sqrt{1 + \beta^2(t_2)}\right)\beta(t_1)}{\left(1 + \sqrt{1 + \beta^2(t_1)}\right)\beta(t_2)}\right| = L_0.$$

From Eq. (4.12) by simple calculation (see appendix **C** Eqs.(C.4-C.6)) one obtain

$$\beta(t_2) = \frac{\beta(t_1)}{\cosh\left(\frac{aL_0}{c^2}\right) + \sinh\left(\frac{aL_0}{c^2}\right)\sqrt{1 + \beta^2(t_1)}} \qquad (4.13)$$

Substitution Eqs.(4.13), into Eq.(4.4) gives

$$l_{\text{ph}}(t) = sgn(t_2 - t_1)\frac{c^2}{a} \int_{t_1=t}^{t_2(t)} \frac{dt}{t} =$$

$$= \frac{c^2}{a} \ln\left[\cosh\left(\frac{aL_0}{c^2}\right) + \sinh\left(\frac{aL_0}{c^2}\right)\sqrt{1 + \beta^2(t_1)}\right]. \qquad (4.14)$$

Suppose that $\frac{aL_0}{c^2} \gg 1$. From Eq. (4.12) by simple calculation (see appendix **C** Eqs.(C.1)-(C.3)) one obtain

$$\beta(t_2) = \frac{2\cdot\beta(t_1)}{1+\sqrt{1+\beta^2(t_1)}} \exp\left(-\frac{aL_0}{c^2}\right) \qquad (4.13')$$

Substitution Eqs.(4.13′), into Eq.(4.4) gives

$$l_{\mathbf{ph}}(t) = sgn(t_2-t_1)\frac{c^2}{a}\int_{t_1=t}^{t_2(t)} \frac{dt}{t} =$$

$$= \frac{c^2}{a}\ln\left(\frac{1}{2}\exp\left(\frac{aL_0}{c^2}\right)+\frac{1}{2}\exp\left(\frac{aL_0}{c^2}\right)\sqrt{1+\beta^2(t)}\right) \qquad (4.14')$$

$$\propto L_0 + \frac{c^2}{a}\ln\left(\frac{at}{c}\right), t\to\infty,$$

$$\beta(t) = \frac{at}{c}.$$

Eq.(4.14) originally was obtained S.A.Podosenov by the next clear geometrical consideration: Let us consider nonlinear transformations from Minkowski frame (1.5) to an noninertial FR.

$$x_I^\mu = x^\mu = \Psi^\mu(y^k,\xi^0)$$

$$\mu = 0,1,2,3,4 \qquad (4.15)$$

$$k = 1,2,3$$

From Eq.(4.15) one obtain

$$dx_I^\mu = dx^\mu = d\Psi^\mu(y^k,\xi^0) =$$

$$= \frac{\partial\Psi^\mu}{\partial y^k}dy^k + \frac{\partial\Psi^\mu}{\xi^0}d\xi^0. \qquad (4.16)$$

Substitution Eq.(1.16) into Eq.(1.5) gives

$$ds^2 = \left(dy^{\hat{0}}\right)^2 + g^*_{\mu\nu}\frac{\partial \Psi^\mu}{\partial y^{\hat{n}}}\frac{\partial \Psi^\nu}{\partial y^{\hat{k}}}dy^{\hat{n}}dy^{\hat{k}} =$$

$$= \left(dy^{\hat{0}}\right)^2 + dl^2,$$

$$g^*_{\mu\nu} = g_{\mu\nu} - V_\mu V_\nu, \tag{4.17}$$

$$V^\mu = \frac{dx^\mu}{d\xi^{\hat{0}}},$$

$$dy^{\hat{0}} = d\xi^{\hat{0}} + V_\mu \frac{\partial \Psi^\mu}{\partial y^{\hat{n}}}dy^{\hat{n}} = V_\mu dx^\mu.$$

Using Eqs.(4.17) one obtain

$$V^{\hat{k}} = \frac{dy^{\hat{k}}}{d\xi^{\hat{0}}} = 0,$$

$$V_{\hat{k}} = V_\mu \frac{\partial \Psi^\mu}{\partial y^{\hat{k}}} = g_{\hat{k}\hat{a}}V^{\hat{a}} = g_{\hat{k}\hat{0}}V^{\hat{0}} = \tag{4.18}$$

$$= \frac{g_{\hat{k}\hat{0}}}{\sqrt{g_{\hat{0}\hat{0}}}}.$$

Hence

$$\left(d\hat{l}\right)^2 = \left(\frac{g_{\hat{n}\hat{0}}g_{\hat{k}\hat{0}}}{g_{\hat{0}\hat{0}}} - g_{\hat{n}\hat{k}}\right)dy^{\hat{n}}dy^{\hat{k}}. \tag{4.19}$$

Using orthogonality condition (Pfaff equation)

$$V_\mu dx^\mu = 0, \mu = 0, 1, 2, 3. \qquad (4.20)$$

by substitution Eq.(4.5) into Eq.(4.20) we obtain:

$$\frac{dx^1}{dx^0} = \frac{dx^1}{dt} = \frac{c\sqrt{1+\beta^2(t)}}{\beta(t)},$$

$$\beta(t) = \frac{at}{c}. \qquad (4.21)$$

Hence

$$x^1(t) = \frac{c^2}{a}\left(\sqrt{1+\beta^2(t)} - \ln\left|\frac{1+\sqrt{1+\beta^2(t)}}{\beta(t)}\right|\right) + A. \qquad (4.22)$$

Setting $t = t_2$ and take into account boundary condition $x^1(t_2) = x^1(y^1, t_2)(y^1 = 0)$ and corresponding with J.S. Bell's problem Eq.(4.23) (see Eq.(4.5))

$$x^1_\mathbf{C}(y^1, t) = x^1(y^1, t) = y^1_\mathbf{C} + \frac{c^2}{a}\left(\sqrt{1+\beta^2(t)} - 1\right)$$

$$x^1_\mathbf{B}(y^1, t) = x^1(y^1, t) = y^1_\mathbf{B} + \frac{c^2}{a}\left(\sqrt{1+\beta^2(t)} - 1\right) \qquad (4.23)$$

$$y^1_\mathbf{C} = 0, y^1_\mathbf{B} = L_0,$$

one obtain

$$\frac{c^2}{a}\left(\sqrt{1+\beta^2(t_2)} - \ln\left|\frac{1+\sqrt{1+\beta^2(t_2)}}{\beta(t_2)}\right|\right) + A(t_2) = \frac{c^2}{a}\left(\sqrt{1+\beta^2(t_2)} - 1\right),$$

<div align="center">hence</div> (4.24)

$$-\frac{c^2}{a}\ln\left|\frac{1+\sqrt{1+\beta^2(t_2)}}{\beta(t_2)}\right| + A(t_2) = -\frac{c^2}{a}.$$

Hence

$$A(t_2) = \frac{c^2}{a}\ln\left|\frac{1+\sqrt{1+\beta^2(t_2)}}{\beta(t_2)}\right| - \frac{c^2}{a}. \qquad (4.26)$$

Setting $t = t_1$ and take into account boundary condition $x^1(t_1) = x_B^1(y^1 = L_0, t_1)$ and corresponding (with J.S. Bell's problem) Eq.(4.23) one obtain

$$\frac{c^2}{a}\left(\sqrt{1+\beta^2(t_1)} - \ln\left|\frac{1+\sqrt{1+\beta^2(t_1)}}{\beta(t_1)}\right|\right) + A(t_2) =$$

$$= L_0 + \frac{c^2}{a}\left(\sqrt{1+\beta^2(t_1)} - 1\right), \qquad (4.27)$$

<div align="center">hence</div>

$$-\frac{c^2}{a}\ln\left|\frac{1+\sqrt{1+\beta^2(t_1)}}{\beta(t_1)}\right| + A(t_2) = L_0 - \frac{c^2}{a}.$$

Substitution Eq.(4.26) into Eq.(4.27) gives

$$L_0 = \frac{c^2}{a}\ln\left|\frac{\left(1+\sqrt{1+\beta^2(t_2)}\right)\cdot\beta(t_1)}{\left(1+\sqrt{1+\beta^2(t_1)}\right)\cdot\beta(t_2)}\right| \qquad (4.28)$$

By using Eqs.(4.28),(4.13) one obtain

$$L(t) = \text{sgn}(t_2 - t_1) \frac{c^2}{a} \int\limits_{t_1=t}^{t_2(t_1)} \frac{dt}{t} =$$

(4.29)

$$= \frac{c^2}{a} \ln\left[ \cosh\left(\frac{aL_0}{c^2}\right) + \sinh\left(\frac{aL_0}{c^2}\right) \sqrt{1 + \beta^2(t_1)} \right].$$

Eq.(4.29) coincide with Eq.(4.14).

# V. J.S. Bell's problem in canonical parametrization by using proper time $\tau$.

Let us consider Bell's problem in canonical parametrization by using proper time $\tau$ such that ($c = 1$) [8]:

$$x_B^1(\tau) = x_B^1(\tau, x_B^0) = \frac{1}{a}[\cosh(a \cdot \tau) - 1] + x_B^0,$$

$$x_A^1(\tau) = x_A^1(\tau, x_A^0) = \frac{1}{a}[\cosh(a \cdot \tau) - 1] + x_A^0,$$

$$x_B^0 - x_A^0 = L_0,$$

(5.1)

$$\tau(t) = \frac{1}{a} \ln\left(a \cdot t + \sqrt{1 + a^2 \cdot t^2}\right),$$

$$t(\tau) = \frac{1}{a} \sinh(a \cdot \tau),$$

$$\rho = x - \frac{1}{a}[\cosh(a \cdot \tau(t)) - 1].$$

By the canonical way the interval $ds^2$ of the FR comuving both rockets takes the form [8]:

$$ds^2 = d\tau^2 - d\rho^2 - 2d\tau d\rho \sinh(a \cdot \tau) - dy^2 - dz^2.$$

(5.2)

$$g_{\tau\tau} = 1, g_{\rho\rho} = -1, g_{\tau\rho} = -\sinh(a \cdot \tau), g_{yy} = -1, g_{zz} = -1$$

By using Eqs.(4.1) we obtain

$$ds^2 = [d\tau - d\rho \sinh(a \cdot \tau)]^2 - d\rho^2 \cosh(a \cdot \tau) - dy^2 - dz^2.$$

(5.3)

By using Eq.(5.3) and Eq.(4.3) we obtain:

$$\frac{dx^1}{dx^0} = \frac{d\rho}{d\tau} = \frac{1}{\sinh(a \cdot \tau)}.$$

(5.4)

By integration Eq.(5.4) we obtain

$$\rho(\tau) = \frac{1}{a} \ln \left| \tanh\left(\frac{a \cdot \tau}{2}\right) \right| + A.$$

(5.5)

Setting $\tau = \tau_2$ and take into account boundary condition $\rho(\tau_2) = x_A^0 = 0$ one obtain:

$$\frac{1}{a} \ln \left| \tanh\left(\frac{a \cdot \tau_2}{2}\right) \right| + A(\tau_2) = x_A^0.$$

(5.6)

$$\tau_2 = \tau_2(\tau_1).$$

Setting $\tau = \tau_1$ and take into account corresponding boundary condition $\rho(\tau)|_{\tau=\tau_1} = x_B^0 = L_0$ one obtain:

$$\frac{1}{a} \ln \left| \tanh\left(\frac{a \cdot \tau_1}{2}\right) \right| + A(\tau_2) = L_0.$$

(5.7)

By subtracting Eq.(5.7) from Eq.(5.6) one obtain:

$$\frac{1}{a} \ln \left| \frac{\tanh\left(\frac{a \cdot \tau_2}{2}\right)}{\tanh\left(\frac{a \cdot \tau_1}{2}\right)} \right| = -L_0. \tag{5.8}$$

From Eq.(5.8) we obtain

$$\frac{\tanh\left(\frac{a \cdot \tau_2(\tau_1)}{2}\right)}{\tanh\left(\frac{a \cdot \tau_1}{2}\right)} = \exp(-a \cdot L_0),$$

$$\tanh\left(\frac{a \cdot \tau_2(\tau_1)}{2}\right) = \left(\tanh\left(\frac{a \cdot \tau_1}{2}\right)\right) \cdot \exp(-a \cdot L_0). \tag{5.9}$$

Hence

$$\tau_2(\tau_1) = \frac{2}{a} \mathbf{Arth}\left[\left(\tanh\left(\frac{a \cdot \tau_1}{2}\right)\right) \cdot \exp(-a \cdot L_0)\right] \tag{5.10}$$

By using Eqs.(4.4),(5.2),(5.4) we obtain:

$$l_{\mathbf{ph}}(\tau) = sgn(\tau_2 - \tau_1) \int_{\tau_1=\tau}^{\tau_2} \sqrt{\cosh^2(a \cdot \tau) d\rho^2(\tau)} =$$

$$= sgn(\tau_2 - \tau_1) \int_{\tau_1=\tau}^{\tau_2} \left(\cosh(a \cdot \tau) \frac{d\rho(\tau)}{d\tau}\right) d\tau =$$

$$= sgn(\tau_2 - \tau_1) \int_{\tau_1}^{\tau_2} \coth(a \cdot \tau) d\tau =$$

$$= sgn(\tau_2 - \tau_1) \frac{1}{a} \ln \left| \frac{\sinh(a \cdot \tau_2(\tau_1))}{\sinh(a \cdot \tau_1)} \right|. \tag{5.11}$$

By substitution Eq.(5.10) into Eq.(5.11) we obtain:

$$L_{ph}(\tau_0) = sgn(\tau_2(\tau_1) - \tau_1)\frac{1}{a} \ln\left|\frac{\sinh(a \cdot \tau_2(\tau_1))}{\sinh(a \cdot \tau_1)}\right| =$$

$$= sgn(\tau_2(\tau_1) - \tau_1)\frac{1}{a} \ln\left|\frac{\sinh\left[2 \cdot \mathbf{Arth}\left[\left(\tanh\left(\frac{a \cdot \tau_1}{2}\right)\right) \cdot \exp(-a \cdot L_0)\right]\right]}{\sinh(a \cdot \tau_1)}\right|. \quad (5.12)$$

By using equality: $\sinh[2 \cdot \mathbf{Arth}(z)] = \frac{2z}{1-z^2}, (z^2 < 1)$ one obtain

$$\sinh\left[2 \cdot \mathbf{Arth}\left[\left(\tanh\left(\frac{a \cdot \tau_1}{2}\right)\right) \cdot \exp(-a \cdot L_0)\right]\right] =$$

$$= \frac{2\left(\tanh\left(\frac{a \cdot \tau_1}{2}\right)\right) \cdot \exp(-a \cdot L_0)}{1 - \left(\tanh^2\left(\frac{a \cdot \tau_1}{2}\right)\right) \cdot \exp(-2a \cdot L_0)}. \quad (5.13)$$

By substitution Eq.(5.13) into Eq.(5.12) finally we obtain

$$L_{ph}(\tau_1) = -\frac{1}{a} \ln\left[\frac{2\left(\tanh\left(\frac{a \cdot \tau_1}{2}\right)\right) \cdot \exp(-a \cdot L_0)}{\left(1 - \left(\tanh^2\left(\frac{a \cdot \tau_1}{2}\right)\right) \cdot \exp(-2a \cdot L_0)\right) \cdot \sinh(a \cdot \tau_1)}\right]$$

$$L_0 + \frac{1}{a} \ln[\sinh(a \cdot \tau_1)] - \frac{1}{a} \ln\left[2\left(\tanh\left(\frac{a \cdot \tau_1}{2}\right)\right)\right] + \quad (5.14)$$

$$+ \frac{1}{a} \ln\left[\left(1 - \left(\tanh^2\left(\frac{a \cdot \tau_1}{2}\right)\right) \cdot \exp(-2a \cdot L_0)\right)\right].$$

Suppose that: $a \cdot L_0 \gg 1, a \cdot \tau_1 \gg 1$. Thus from Eq.(5.14) we obtain

$$L_{ph}(\tau_1) \propto L_0 + \frac{1}{a} \ln[\sinh(a \cdot \tau_1)] \quad (5.15)$$

## Appendix A.

We denote space-time indices $0, 1, 2, 3$ in Greek (where $0$ corresponds to the time dimension), while spatial indices $1, 2, 3$ are denoted in Roman. We assume that summation takes a place on two same indices met in the same term. We assume that

$$ds^2 = g_{\alpha\beta}dx^\alpha dx^\beta,$$

$$x^0 = ct.$$

(A.1)

Targeting the A.Zelmanov's chronometrically invariant formulae for the elementary "length" [4] $dl$, the it is *a pure spatial metric tensors* $h_{ik}$ and $h^{ik}$, and the fundamental determinant $h = |h_{ik}|$, one obtain

$$dl^2 = h_{ik}dx^i dx^j,$$

$$h_{ik} = -g_{ik} + \frac{g_{ik}g_{0k}}{g_{00}},$$

$$h^{ik} = -g^{ik}, h = -\frac{g}{g_{00}}.$$

(A.2)

where $g = |g_{\mu\nu}|$. The spatial metric determined in such a way coincides with "*radar metric*" that assumed by Landau and Lifshitz, see (84.6) and (84.7) in [5], and that assumed by Fock, see (55.20) in [6]. For the elementary chronometrically invariant interval of time $d\tau$ and the elementary worldinterval $ds$, one obtain

$$cd\tau = \frac{g_{0\alpha}dx^\alpha}{\sqrt{g_{00}}}, ds^2 = c^2 d\tau^2 - dl^2.$$

(A.3)

# Appendix B.

$$2 \cdot \mathbf{Arth}(z) = \ln \frac{1+z}{1-z}, z^2 < 1. \qquad (B.1)$$

$$\sinh[2 \cdot \mathbf{Arth}(z)] = \sinh\left[\ln \frac{1+z}{1-z}\right] =$$

$$= \frac{1}{2}\left[\exp\left(\ln \frac{1+z}{1-z}\right) - \exp\left(-\ln \frac{1+z}{1-z}\right)\right] =$$

$$\frac{1}{2}\left[\exp\left(\ln \frac{1+z}{1-z}\right) - \exp\left(\ln \frac{1-z}{1+z}\right)\right] = \qquad (B.2)$$

$$\frac{1}{2}\left[\frac{1+z}{1-z} - \frac{1-z}{1+z}\right] = \frac{1}{2}\frac{(1+z)^2 - (1-z)^2}{(1-z)(1+z)} =$$

$$\frac{1}{2}\frac{4z}{1-z^2} = \frac{2z}{1-z^2}.$$

# Appendix C.

$$L_0 = \frac{c^2}{a} \ln \left| \frac{\left(1 + \sqrt{1 + \beta^2(t_2)}\right) \cdot \beta(t_1)}{\left(1 + \sqrt{1 + \beta^2(t_1)}\right) \cdot \beta(t_2)} \right| \qquad (C.1)$$

By using Eq.(C.1) one obtain

$$\frac{\left(1 + \sqrt{1 + \beta^2(t_2)}\right) \cdot \beta(t_1)}{\left(1 + \sqrt{1 + \beta^2(t_1)}\right) \cdot \beta(t_2)} = \exp\left(\frac{aL_0}{c^2}\right) \triangleq d \qquad (C.2)$$

Eq.(C.2) gives

$$\frac{1 + \sqrt{1 + \beta^2(t_2)}}{\beta(t_2)} = \frac{d \cdot \left(1 + \sqrt{1 + \beta^2(t_1)}\right)}{\beta(t_1)} \triangleq d_1,$$

$$1 + \sqrt{1 + \beta^2(t_2)} = d_1 \cdot \beta(t_2),$$

$$\sqrt{1 + \beta^2(t_2)} = d_1 \cdot \beta(t_2) - 1,$$

$$1 + \beta^2(t_2) = d_1^2 \cdot \beta^2(t_2) - 2d_1 \cdot \beta(t_2) + 1, \qquad (C.3)$$

$$(d_1^2 - 1) \cdot \beta^2(t_2) - 2d_1 \cdot \beta(t_2) = 0$$

$$\beta(t_2) = \frac{2d_1}{d_1^2 - 1} \simeq \frac{2}{d_1}, \quad \frac{aL_0}{c^2} \gg 1,$$

$$\beta(t_2) = \frac{2 \cdot \beta(t_1)}{1 + \sqrt{1 + \beta^2(t_1)}} \exp\left(-\frac{aL_0}{c^2}\right)$$

$$\frac{d \cdot \left(1 + \sqrt{1 + \beta^2(t_1)}\right)}{\beta(t_1)} \triangleq d_1,$$

$$d_1^2 - 1 = \frac{d^2 \cdot \left(2 + 2\sqrt{1 + \beta^2(t_1)} + \beta^2(t_1)\right)}{\beta^2(t_1)} - 1 =$$

$$= \frac{d^2 \cdot \left(2 + 2\sqrt{1 + \beta^2(t_1)} + \beta^2(t_1)\right) - \beta^2(t_1)}{\beta^2(t_1)} =$$

$$= \frac{2d^2 \cdot \left(1 + \sqrt{1 + \beta^2(t_1)}\right) + \beta^2(t_1)(d^2 - 1)}{\beta^2(t_1)} =$$

$$\frac{d_1}{d_1^2 - 1} = \frac{d \cdot \left(1 + \sqrt{1 + \beta^2(t_1)}\right)}{\beta(t_1)} \cdot \frac{\beta^2(t_1)}{2d^2 \cdot \left(1 + \sqrt{1 + \beta^2(t_1)}\right) + \beta^2(t_1)(d^2 - 1)} =$$

$$= \frac{d \cdot \beta^2(t_1)}{\beta(t_1)\left(\sqrt{1 + \beta^2(t_1)} - 1\right)} \cdot \frac{\beta^2(t_1)}{2d^2 \cdot \left(1 + \sqrt{1 + \beta^2(t_1)}\right) + \beta^2(t_1)(d^2 - 1)} = \quad (C.4)$$

$$\frac{d \cdot \beta(t_1) \cdot \beta^2(t_1)}{2d^2 \cdot \left(1 + \sqrt{1 + \beta^2(t_1)}\right)\left(\sqrt{1 + \beta^2(t_1)} - 1\right) + \beta^2(t_1)(d^2 - 1)\left(\sqrt{1 + \beta^2(t_1)} - 1\right)}$$

$$= \frac{d \cdot \beta(t_1) \cdot \beta^2(t_1)}{2d^2 \cdot \beta^2(t_1) + \beta^2(t_1)(d^2 - 1)\left(\sqrt{1 + \beta^2(t_1)} - 1\right)} =$$

$$= \frac{d \cdot \beta(t_1)}{2d^2 + (d^2 - 1)\left(\sqrt{1 + \beta^2(t_1)} - 1\right)} =$$

$$= \frac{d \cdot \beta(t_1)}{2d^2 + (d^2 - 1)\sqrt{1 + \beta^2(t_1)} - (d^2 - 1)} =$$

$$= \frac{d \cdot \beta(t_1)}{(d^2 + 1) + (d^2 - 1)\sqrt{1 + \beta^2(t_1)}} = \frac{\beta(t_1)}{\left(d + \frac{1}{d}\right) + \left(d - \frac{1}{d}\right)\sqrt{1 + \beta^2(t_1)}}.$$

Hence

$$\beta(t_2) = \frac{2d_1}{d_1^2 - 1} = \frac{2\beta(t_1)}{\left(d + \frac{1}{d}\right) + \left(d - \frac{1}{d}\right)\sqrt{1 + \beta^2(t_1)}} =$$

$$= \frac{\beta(t_1)}{\frac{1}{2}\left(d + \frac{1}{d}\right) + \frac{1}{2}\left(d - \frac{1}{d}\right)\sqrt{1 + \beta^2(t_1)}} = \qquad (C.5)$$

$$\frac{\beta(t_1)}{\frac{1}{2}\left(\exp\left(\frac{aL_0}{c^2}\right) + \exp\left(-\frac{aL_0}{c^2}\right)\right) + \frac{1}{2}\left(\exp\left(\frac{aL_0}{c^2}\right) - \exp\left(-\frac{aL_0}{c^2}\right)\right)\sqrt{1 + \beta^2(t_1)}} =$$

$$= \frac{\beta(t_1)}{\cosh\left(\frac{aL_0}{c^2}\right) + \sinh\left(\frac{aL_0}{c^2}\right)\sqrt{1 + \beta^2(t_1)}}.$$

Finally we obtain

$$\beta(t_2) = \frac{\beta(t_1)}{\cosh\left(\frac{aL_0}{c^2}\right) + \sinh\left(\frac{aL_0}{c^2}\right)\sqrt{1 + \beta^2(t_1)}}. \qquad (C.6)$$